\documentclass[twocolumn,aps,showpacs]{revtex4}
\usepackage{epsf,graphicx}
\usepackage{supercite}
\usepackage{subfigure}
\begin{document}
\title{Thermodynamical Properties of a Rotating Ideal Bose Gas}
\author{Sebastian Kling}
\email{kling@iap.uni-bonn.de}
\affiliation{%
Institut f{\"u}r Angewandte Physik, Universit{\"a}t Bonn, Wegelerstra{\ss}e 8,
53115 Bonn, Germany}
\author{Axel Pelster}
\email{axel.pelster@uni-due.de}
\affiliation{%
Fachbereich Physik, Campus Duisburg, 
Universit{\"a}t Duisburg-Essen, Lotharstrasse 1,
47048 Duisburg, Germany}      

\date{\today}

\begin{abstract}
In a recent experiment, a Bose-Einstein condensate was trapped in an anharmonic potential which is well 
approximated by a harmonic and a quartic part. 
The condensate was set into such a fast rotation that the centrifugal force in the corotating frame 
overcompensates the harmonic part in the plane perpendicular to the rotation axis. 
Thus, the resulting trap potential became Mexican-hat shaped. 
We present an analysis for an ideal Bose gas which is confined in such an anharmonic rotating trap within 
a semiclassical approximation where we calculate the critical temperature, the condensate fraction, and the 
heat capacity. 
In particular, we examine in detail how these thermodynamical quantities depend on the rotation frequency. 
\end{abstract}
\maketitle
%
\section{Introduction}
\begin{figure}[t]
\center\begin{minipage}[tb]{.4\textwidth}
\center\includegraphics[scale=.65]{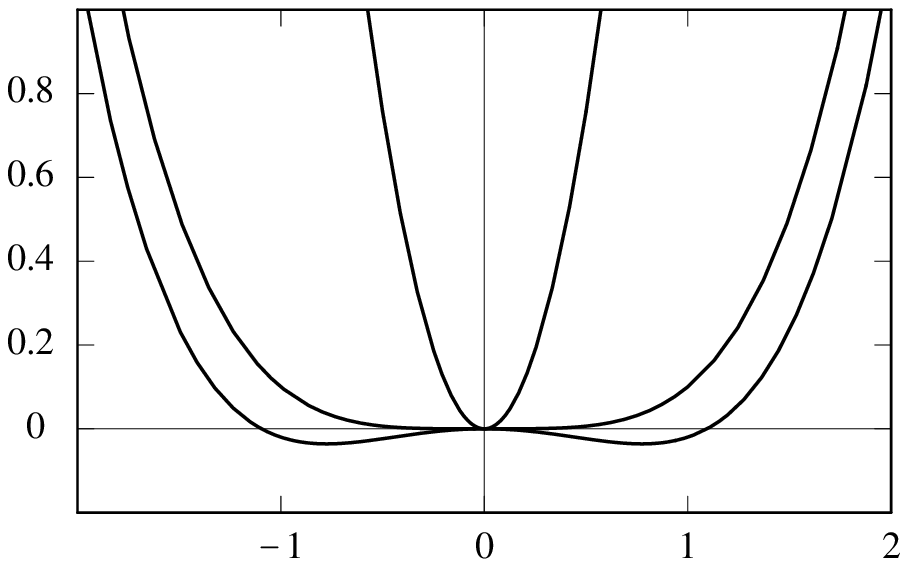}
\put(-3.5,-.3){$x\left[\frac{\hbar}{M\omega_z}\right]$}
\put(-7,1){\rotatebox{90}{$V(x,0,0)\,\,\left[\hbar\omega_z\right]$}}
\put(-3.3,2.3){$\Omega_1$}
\put(-2,2){$\Omega_2$}
\put(-1.5,1.1){$\Omega_3$}
\renewcommand{\figurename}{FIG.}
\caption{Trapping potential (\ref{Vrot1}) in $(x,0,0)$-direction for the values of the Paris experiment \cite{Bretin} 
and varying rotation frequencies $\Omega_1=0$, $\Omega_2=\omega_\bot$, $\Omega_3=1.04\times\omega_\bot$. The latter is the largest experimentally realized 
rotation frequency.}
\label{POT}
\end{minipage}
\end{figure}

The rotation of a quantum fluid leads to many interesting problems. In particular, the rotation of ultracold
Bose gases has motivated a lot of scientific work within the last few years.\\
The rotation has the effect of adding an angular momentum to the system so that it
can be compared with the motion of a charged particle in a magnetic field. 
Initial experiments \cite{Dali00,Ketterle01} have shown that, analogously to the superfluid
helium II \cite{Donelly}, vortices 
nucleate at some critical rotation frequency. 
The main difference between earlier studies about quantum fluids and BEC's is the trap in which the condensate is 
confined. 
This inspired A.L.~Fetter \cite{Fetter01} to suggest adding a quartic term to the harmonic trap potential. 
His idea was to rotate the condensate so fast that the centrifugal force may overcompensate the harmonic trapping 
potential.
In a harmonic trap, the condensate gets lost when the rotation frequency comes close to the harmonic trap 
frequency \cite{Rosenbusch}, but the additional anharmonicity ensures the confining of the condensate. 
Such a trap was realized in an experiment in Paris at the {\'E}cole Normale Sup{\'e}rieure (ENS) in the group of J.~Dalibard 
in 2004 \cite{Bretin}. \\
The experimental setup was a usual magneto-optical trap with the frequencies $\omega_x=\omega_y=2\pi\times75.5$~Hz and $\omega_z=2\pi\times 11.0$~Hz.
The additional quartic anharmonicity was generated with a Gaussian laser beam propagating in $z$-direction which created 
a potential $U= U_0\exp{(-2r_\bot^2/w^2)}$ with the perpendicular radius $r_\bot= \sqrt{x^2+y^2}$, the laser's waste 
$w=25$~$\mu$m,  and the intensity $U_0\sim(k_B\times90)$~nK. 
Due to the experimentally realized condition $r_\bot<w/2$, this potential is well approximated by 
$U=U_0-(2U_0/w^2)r_\bot^2+(2U_0/w^4)r_\bot^4$. 
In Ref.~\cite{Bretin} an amount of $3\times 10^5$ atoms of ${}^{87}$Rb is set into rotation with another 
laser beam acting as a stirrer. 
The laser creates an anisotropic potential in the $xy$-plane which rotates with frequency $\Omega$.
In the corotating frame, the resulting trapping potential  can be written as
\begin{equation}
\label{Vrot1}
V_{\rm rot}({\bf x},\Omega)=\frac{M}{2}\left(\omega_\bot^2-\Omega^2\right)r_\bot^2+\frac{M}{2}\omega_z^2z^2+\frac{k}{4}r_\bot^4\,,
\end{equation}
where $\omega_\bot=\omega_x-4U_0M^{-1}w^{-2}=2\pi\times64.8$ Hz and $M$ is the atomic mass.
The last term in (\ref{Vrot1}) corresponds to the quartic anharmonicity with $k=8U_0 w^{-4}=2.6\times10^{-11}$~Jm${}^{-4}$.\\
In the following we treat the Bose gas in the anharmonic trap (\ref{Vrot1}) within the grand-canonical ensemble 
and determine the critical temperature, the condensate fraction, and the heat capacity of the Bose gas within a 
semiclassical approximation. 
In our discussion the rotation frequency $\Omega$ appears as a control parameter. 
The Paris experiment \cite{Bretin} allows rotation frequencies $\Omega$ up to $1.04\times\omega_\bot$. 
However, in the present theoretical discussion, we consider arbitrarily large rotation frequencies. 
In Fig.~\ref{POT} we depict how the trapping potential (\ref{Vrot1}) varies with increasing rotation frequency $\Omega$. 
For small rotation frequencies $\Omega<\omega_\bot$, the potential (\ref{Vrot1}) is convex, for the critical rotation frequency 
$\Omega=\omega_\bot$ it is purely quartic in the perpendicular plane, and for a fast rotation frequency $\Omega>\omega_\bot$ the trap has the 
shape of a Mexican hat. 
In our discussion, we focus on two particular rotation frequencies, namely the critical rotation frequency $\Omega=\omega_\bot$ 
and the limit of an infinite fast rotation frequency $\Omega\to\infty$ as the potential reduces to power laws in both cases.
Such potentials were investigated some time ago as they lead to analytic formulas for the respective thermodynamical 
properties \cite{Bagnato87,Yukalov}. 
Note that the case $\Omega\to\infty$ corresponds to a trap where the bosons are confined to a cylinder of radius
$r_{\rm cyl}=\sqrt{M^2\omega_z(\Omega^2-\omega_\bot^2)/(k\hbar)}$. 
Thus, the above mentioned experimental restriction $r_{\rm cyl}<w/2$ allows to determine a maximum rotation frequency 
$\Omega_{\rm max}$ for which the anharmonic potential (\ref{Vrot1}) of the Paris experiment is valid. 
The resulting value $\Omega_{\rm max}=1.08\times\omega_\bot$ shows that considering an infinite rotation frequency $\Omega\to\infty$ is not 
suitable for this experiment. 
\section{Ideal Bose Gas in rotating Trap}
\setlength{\unitlength}{1cm}
We consider $N$ particles of an ideal Bose gas which are distributed over various quantum states $\nu$ of the system. 
These states are characterized through the population $n_{\bf n}$ of the one-particle state ${\bf n}$ of the trap 
(\ref{Vrot1}), such that the energy levels are $E_\nu = \sum_{\bf n}n_{\bf n}E_{\bf n}$, where $E_{\bf n}$ denotes the 
one-particle energy. 
Correspondingly, the number of particles in state $\nu$ are given by $N_\nu = \sum_{\bf n}n_{\bf n}$. 
The resulting grand-canonical ensemble is determined by its partition function
\begin{equation}
\label{ZG1}
\mathcal{Z} = \sum_{\nu}\exp{\left[-\beta\left(E_\nu-\mu N_\nu\right)\right]}\,,
\end{equation}
where $\beta=1/(k_B T)$ denotes an inverse temperature, $k_B$ is Boltzmann's constant, and $\mu$ is the chemical potential. 
The corresponding grand-canonical free energy $\mathcal{F} = -(1/ \beta)\ln{\mathcal{Z}}$ allows to calculate all relevant 
thermodynamical quantities \cite{Pitaevskii,Pethick}. 
Around the minimum, the trap (\ref{Vrot1}) has a small curvature so that the energy levels are close to each other. 
With increasing rotation frequency, the curvature decreases until the critical rotation frequency $\Omega=\omega_\bot$ is reached. 
Not until the rotation frequency overcompensates the harmonic part of the trap, the curvature increases again.
Thus, for all experimentally realized rotation frequencies $0\leq\Omega\leq1.04\times\omega_\bot$, our system can be 
described by  the discrete ground state $E_0$, which must be retained quantum-mechanically, plus a continuum of states 
above $E_0$. 
Within this semiclassical approximation, we can set the ground-state energy $E_0$ to zero so that the grand-canonical 
free energy of the ideal Bose gas reads 
\begin{eqnarray}
\label{FG1}
\mathcal{F}&=&N_0(\mu_c-\mu)\nonumber\\
&&\hspace{-8mm}-\sum_{j=1}^\infty\frac{1}{\beta j}\int\frac{d^3\!x\,d^3\!p}{(2\pi\hbar)^3}\,
\exp{\left\{ -\beta j\left[H({\bf x}, {\bf p})-\mu\right]\right\}}\,,
\end{eqnarray} 
where the energy levels are replaced by the classical Hamiltonian
\begin{equation}
H({\bf x}, {\bf p})=\frac{{\bf p}^2}{2M} + V_{\rm rot}({\bf x}, \Omega)\,.
\end{equation}
The critical chemical potential $\mu_c$, where the condensation emerges, is determined by the condition 
$H({\bf x},{\bf p})-\mu_c>0$ and is therefore given by $\mu_c=\min_{\bf x}{V_{\rm rot}({\bf x},\Omega)}$. 
Due to (\ref{Vrot1}) it reads explicitly
\setlength{\unitlength}{1cm}
\begin{figure}[t]
\center\begin{minipage}[tb]{.4\textwidth}
\center\includegraphics[scale=.65]{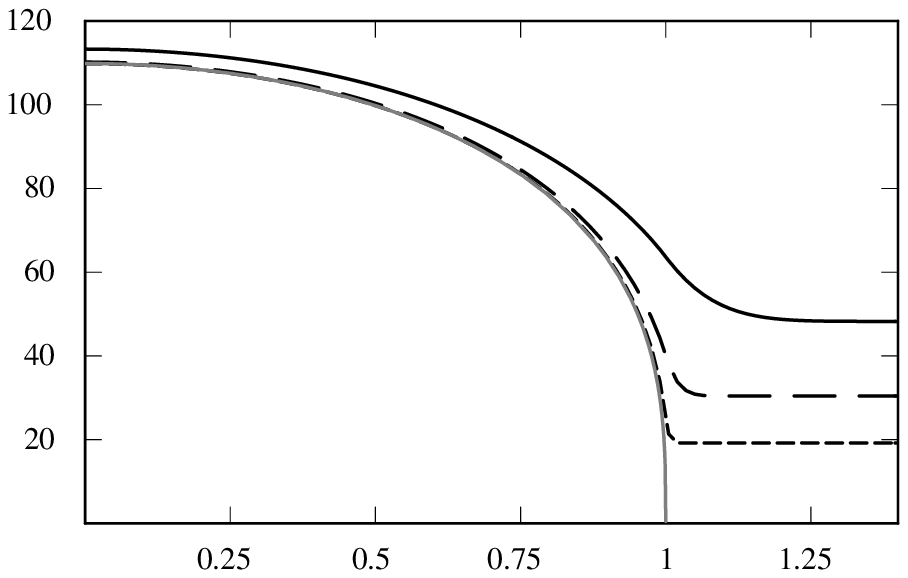}
\put(-4,-.3){$\Omega/ \omega_\bot$}
\put(-7,1.6){\rotatebox{90}{$T_c(\Omega)$  [nK]}}
\renewcommand{\figurename}{FIG.}
\caption[Critical temperature (anharmonic) versus rotation frequency.]{\small Critical temperature versus rotation 
frequency $\Omega$. The highest lying line (solid) corresponds to the data of the Paris experiment \cite{Bretin}, see 
data below Eq.~(\ref{Vrot1}). For the deeper lying lines we varied the anharmonicity parameter $k$: $k\to k/10$ 
(long dashes), $k\to k/100$ (short dashes), and the harmonic limit $k\downarrow0$ (gray solid).}
\label{PTc1}
\end{minipage}
\end{figure}
\begin{equation}
\label{Mu1}
\mu_c=\left\{\begin{array}{cl}
0&;\,\Omega\leq\omega_\bot\,,\\[2mm]
-{\displaystyle \frac{M^2}{4k}\left(\omega_\bot^2-\Omega^2\right)^2}&;\,\Omega>\omega_\bot\,.
\end{array}\right.
\end{equation}
Performing the phase-space integral in (\ref{FG1}), we obtain 
\begin{equation}
\label{FG2}
\mathcal{F}=N_0(\mu_c-\mu)-\frac{\zeta_4(e^{\beta\mu},\Omega)}{\beta^4\hbar^3\omega_z\left(\omega_\bot^2-\Omega^2\right)}\,,
\end{equation}
where we have introduced the generalized $\zeta$-function
\begin{eqnarray}
\label{Zeta1}
\zeta_\nu(e^{\beta\mu},\Omega)&=&\sum_{j=1}^\infty\frac{e^{j\beta\mu}}{j^{\nu}}
\sqrt{j\pi \gamma_T}\left(\omega_\bot^2-\Omega^2\right)\nonumber\\
&&\hspace{-20mm}\times\exp{\left[ j \gamma_T \left(\omega_\bot^2-\Omega^2\right)^2\right]}
\mathrm{erfc}\left[\sqrt{j \gamma_T} \left(\omega_\bot^2-\Omega^2\right)\right]
\end{eqnarray}
with the complementary error function
\begin{equation}
\label{error}
\mathrm{erfc}(z)=\frac{2}{\sqrt{\pi}}\int_z^{\infty}\!\!dt\,e^{-t^2}\,.
\end{equation}
We have also shortened the notation with using $\gamma_T=M^2/(4kk_BT)$ as another inverse temperature.
We remark that in the limit of a vanishing anharmonicity $k\downarrow0$ with an undercritical rotation frequency $\Omega<\omega_\bot$ the 
generalized $\zeta$-function (\ref{Zeta1}) reduces to the polylogarithmic function
\begin{equation}
\label{Zeta2}
\lim_{k\downarrow0 \atop \,\,\,\Omega<\omega_\bot}{\zeta_\nu(z,\Omega)}=\zeta_\nu(z)=\sum_{j=1}^\infty\frac{z^j}{j^{\nu}}\,,
\end{equation}
which is related to the Riemann $\zeta$-function via
\begin{equation}
\label{Zeta3}
\zeta_\nu(1) = \zeta(\nu)=\sum_{j=1}^{\infty}\frac{1}{j^\nu}\,.
\end{equation}
Furthermore, we note that in the limit of the critical rotation frequency $\Omega\to\omega_\bot$, the generalized $\zeta$-function 
(\ref{Zeta1}) reads
\begin{equation}
\label{Zeta5}
\lim_{\Omega\to\omega_\bot}{\frac{\zeta_\nu(z,\Omega)}{\omega_\bot^2-\Omega^2}} = \sqrt{\pi \gamma_T}\zeta_{\nu-1/2}(z)
\end{equation}
and in the limit of an infinite fast rotation it is approximated by
\begin{equation} 
\label{Zeta6} 
\frac{ \zeta_\nu(z,\Omega) }{ \omega_\bot^2-\Omega^2 } \approx 2\sqrt{\pi \gamma_T}\sum_{j=1}^\infty \frac{ z^j 
e^{ j\gamma_T\left(\omega_\bot^2-\Omega^2\right)^2} }{ j^{\nu-1/2} }\,;\, \Omega\to\infty \,, 
\end{equation} 
\section{Condensate Density}
\setlength{\unitlength}{1cm}
\begin{figure}[t]
\center\begin{minipage}[t]{.4\textwidth}
\center\includegraphics[scale=.65]{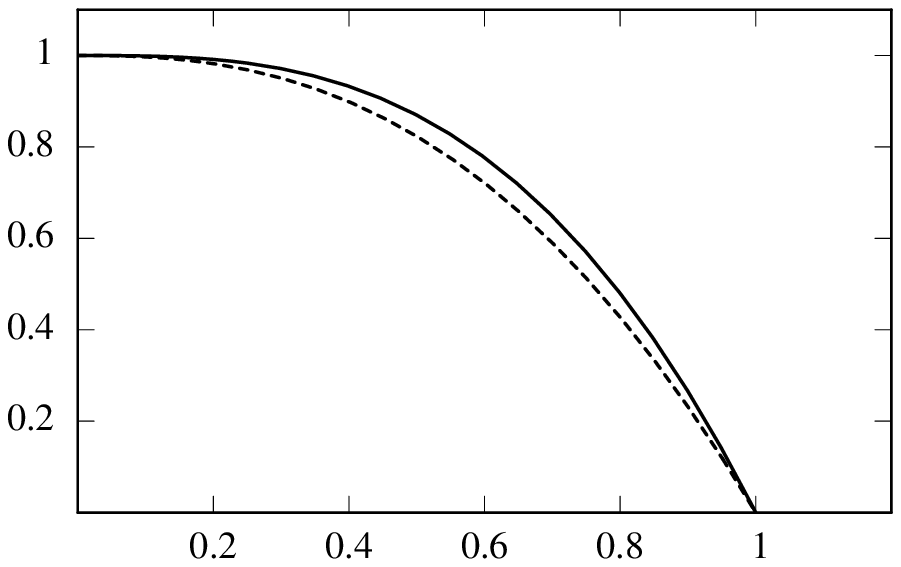}
\put(-7,1.7){\rotatebox{90}{\small $N_0/N$}}
\put(-4,-0.3){\small $T/T_c(\Omega)$}
\renewcommand{\figurename}{FIG.}
\caption{\small Condensate fraction versus reduced temperature. The solid line corresponds to the condensate 
fraction (\ref{CF2}) of a Bose gas in the trap (\ref{Vrot1}) for the rotation frequency $\Omega=0$ and the parameters 
of the Paris experiment \cite{Bretin}. The dashed line corresponds to the condensate fraction at the critical 
$\Omega=\omega_\bot$ and at the infinite fast $\Omega\to\infty$ rotation frequency given by Eq.~(\ref{CF1}).}
\label{PCF1}
\end{minipage}
\end{figure}
From the grand-canonical free energy (\ref{FG2}) we read off that the number of particles 
$N = -(\partial \mathcal{F}/ \partial\mu)_{T,V}$ of an ideal Bose gas is given by a sum of $N= N_0 +N_e$ of particles 
in the ground state $N_0$ and particles in excited states $N_e$:  
\begin{equation}
\label{N1}
N=N_0+\frac{\zeta_3(e^{\beta\mu},\Omega)}{\beta^3\hbar^3\omega_z\left(\omega_\bot^2-\Omega^2\right)}\,.
\end{equation}
The critical temperature $T_c$ at which the condensation emerges can be found from Eq.~(\ref{N1}) by setting $N_0=0$ 
and $\mu=\mu_c$. 
For undercritical rotation frequencies $\Omega<\omega_\bot$ and vanishing anharmonicity $k\downarrow0$, we apply (\ref{Zeta2}) so that the 
critical temperature reads
\begin{equation}
\label{Tc0}
T_c=\frac{\hbar\omega_z}{k_B}\left[\frac{(\omega_\bot^2-\Omega^2)N}{\omega_z^2\zeta(3)}\right]^{1/3}\,;\,\,k=0\,.
\end{equation}
For a non-vanishing anharmonicity $k$, the critical temperature can not be determined explicitly because it appears in 
Eq.~(\ref{N1}) transcendentally.
However, there are two special cases in which we obtain an analytical expression for the critical temperature. 
At first, for the critical rotation $\Omega=\omega_\bot$, we find with (\ref{Zeta5}) \cite{StockLP}: 
\setlength{\unitlength}{1cm}
\begin{figure}[t]
\center\begin{minipage}[t]{.4\textwidth}
\center\includegraphics[scale=.65]{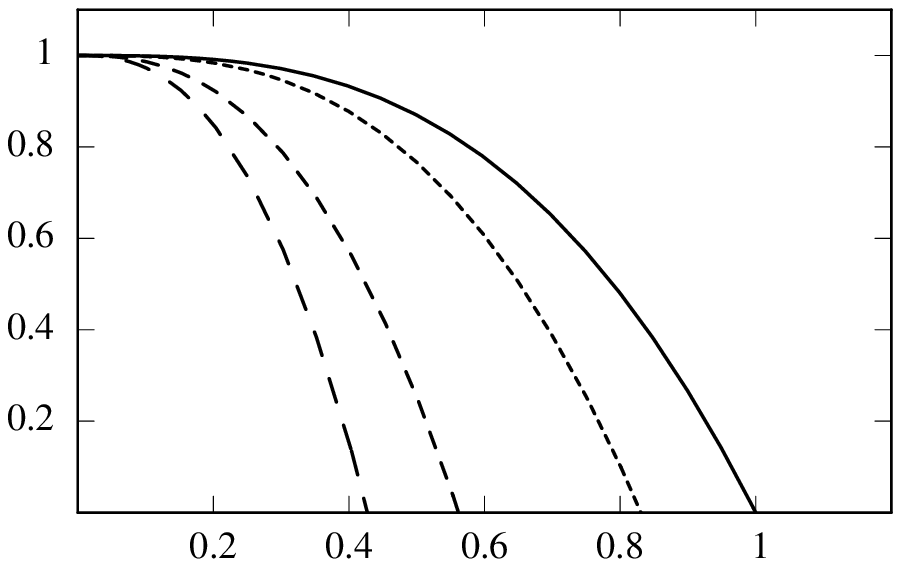}
\put(-7,1.7){\rotatebox{90}{\small $N_0/N$}}
\put(-4,-0.3){\small $T/T_c(\Omega=0)$}
\renewcommand{\figurename}{FIG.}
\caption{\small Condensate fraction versus reduced temperature. The condensate fraction (\ref{CF2}) is evaluated for 
various rotation frequencies $\Omega$, the solid line corresponds to $\Omega=0$ to which the temperature is normalized. The 
other lines are for $\Omega=\omega_\bot/ \sqrt{2}$ (short dashes), $\Omega=\omega_\bot$ (longer dashes), and $\Omega=\sqrt{3/2}\times \omega_\bot$ (long dashes).}
\label{PCF2}
\end{minipage}
\end{figure}
\begin{equation}
\label{Tc1}
T_c = \frac{\hbar\omega_z}{k_B}\left(\frac{4k\hbar}{\pi M^2\omega_z^3}\right)^{1/5} \left(\frac{N}{\zeta(5/2)}\right)^{2/5}\,.
\end{equation}
Secondly, the limit of an infinite fast rotation frequency $\Omega\to\infty$ leads with (\ref{Zeta6}) to the critical temperature
\begin{equation}
\label{Tc2}
T_c = \frac{\hbar\omega_z}{k_B}\left(\frac{k \hbar}{\pi M^2\omega_z^3}\right)^{1/5} \left(\frac{N}{\zeta(5/2)}\right)^{2/5}\,,
\end{equation}
which is by a factor $(1/4)^{1/5} \approx 0.76$ smaller than the previous one. 
A numerical evaluation of the critical temperature obtained from (\ref{N1}) is shown in Fig.~\ref{PTc1} for the 
values of the Paris experiment \cite{StockPhD}. 
For the non-rotating trap, we see that the anharmonicity only slightly 
affects the critical temperature.
With increasing rotation frequency  $\Omega$, the critical temperature decreases and the difference 
between the harmonic and the anharmonic trap is clearly seen. 
At the critical rotation frequency $\Omega=\omega_\bot$, the critical temperature is $T_c=63.5$~nK which is about three times 
smaller than the one estimated for the Paris experiment \cite{Bretin}. \\
From the number of particles (\ref{N1}), we also obtain the condensate fraction in the temperature regime $T<T_c$. 
Here the chemical potential coincides with the critical one given by Eq.~(\ref{Mu1}). 
For undercritical rotation frequencies $\Omega<\omega_\bot$ and vanishing anharmonicity $k\downarrow0$, we use (\ref{Zeta2}) so that the 
condensate fraction is given by
\begin{equation}
\label{CF0}
\frac{N_0}{N}=1-\left(\frac{T}{T_c}\right)^{3}\,;\,k=0\,.
\end{equation}
Furthermore, applying (\ref{Zeta5}) and (\ref{Zeta6}) in the cases $\Omega=\omega_\bot$ and $\Omega\to\infty$, 
respectively, yields with the critical chemical potential (\ref{Mu1}) the following condensate fraction:
\begin{equation}
\label{CF1}
\frac{N_0}{N}=1-\left(\frac{T}{T_c}\right)^{5/2}\,.
\end{equation}
Here, $T_c$ is given by (\ref{Tc0}) and (\ref{Tc1}), (\ref{Tc2}), respectively.
In general, the condensate fraction is given by
\begin{equation}
\label{CF2}
\frac{N_0}{N}=1-\left(\frac{T}{T_c}\right)^{3}\frac{\zeta_3(e^{\beta\mu_c},\Omega)}{\zeta_3(e^{\beta_c\mu_c},\Omega)}\,,
\end{equation}
where $\beta_c=1/(k_BT_c)$. 
In the low-temperature limit $T\!\downarrow\!0$, 
the condensate fraction shows a power-law behavior which is different in the two regimes of undercritical rotation 
$\Omega<\omega_\bot$ and overcritical rotation $\Omega>\omega_\bot$. 
Due to (\ref{Zeta2}) and (\ref{Zeta6}), we obtain
\begin{equation}
\label{CF3}
\hspace{-5mm}\frac{N_0}{N} \approx
\left\{ \begin{array}{ll}
{\displaystyle 1-\frac{k_B^3\zeta(3)}{N\hbar^3\omega_z(\omega_\bot^2-\Omega^2)}\,T^3}&\!\!;\, \Omega<\omega_\bot\,,\\[5mm]
{\displaystyle 1-\frac{M\sqrt{\pi}k_B^{5/2}\zeta(5/2)}{N\sqrt{k}\hbar^3\omega_z}\,T^{5/2} } &\!\!;\, \Omega > \omega_\bot\,.
\end{array}\right. 
\end{equation}
The temperature dependence of the condensate fraction $N_0/N$ following from (\ref{CF2}) is shown in Fig.~\ref{PCF1} 
and Fig.~\ref{PCF2}. 
From this we read off that the temperature dependence of the condensate fraction depends crucially on the rotation 
frequency and is thus not universal. 
\section{Heat Capacity}
The heat capacity follows from the grand-canonical free energy $\mathcal{F}=U-TS-\mu N$, where $U$ is the internal energy 
and $S$ is the entropy, according to
\begin{equation}
\label{HC1}
C=\left.\frac{\partial U}{\partial T}\right|_{N,V}\,.
\end{equation}
Within the grand-canonical ensemble, the heat capacity has to be determined separately in the two regimes $T>T_c$ 
and $T<T_c$.
\subsection{Gase Phase}
At first, we treat the gas phase where $N_0=0$ and determine the entropy via the thermodynamical relation 
$S=-(\partial \mathcal{F}/\partial T)_{V,\mu}$:
\begin{eqnarray}
\label{S1}
\frac{S_>}{k_BN}&=&\frac{7\zeta_4(e^{\beta\mu},\Omega)}{2\zeta_3(e^{\beta\mu},\Omega)}+\gamma_T(\omega_\bot^2-\Omega^2)^2\frac{\zeta_3(e^{\beta\mu})}{\zeta_3(e^{\beta\mu},\Omega)}\nonumber\\[2mm]
&&-\left[\beta\mu+\gamma_T(\omega_\bot^2-\Omega^2)^2\right]
\end{eqnarray}
The internal energy $U=\mathcal{F}+TS+\mu N$ then follows from (\ref{FG2}), (\ref{N1}), (\ref{S1}) and reads
\begin{eqnarray}
\label{U1}
\frac{U_>}{Nk_BT}&=&\frac{5\zeta_4(e^{\beta\mu},\Omega)}{2\zeta_3(e^{\beta\mu},\Omega)}\nonumber\\
&&+\gamma_T(\omega_\bot^2-\Omega^2)^2\left[\frac{\zeta_3(e^{\beta\mu})}{\zeta_3(e^{\beta\mu},\Omega)}-1\right]\,.
\end{eqnarray}
Finally, the heat capacity (\ref{HC1}) for temperatures above $T_c$ is given by
\begin{figure}[t] 
\center\begin{minipage}[t]{.4\textwidth} 
\center\includegraphics[scale=.65]{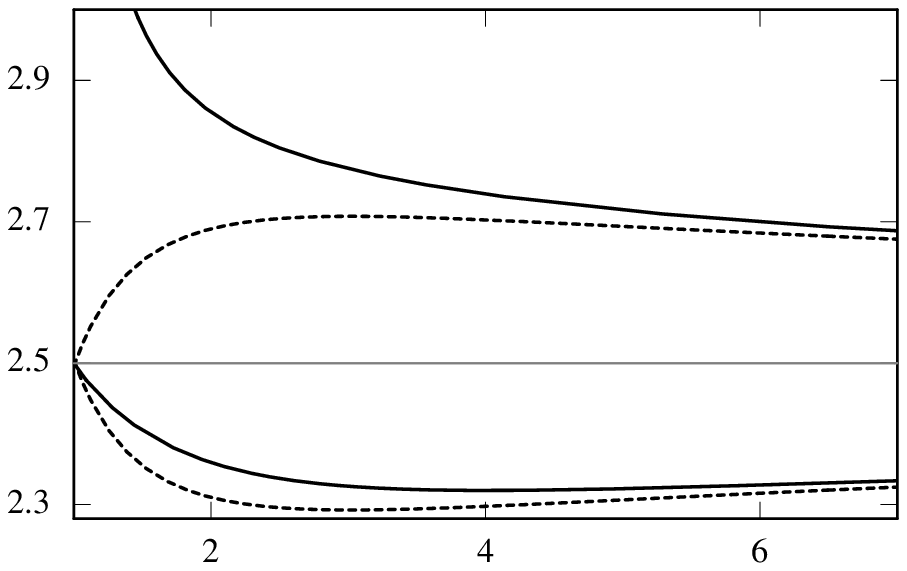} 
\put(-7,1.5){\rotatebox{90}{\small $C/k_BN$}} 
\put(-4,-0.3){\small $T/T_c(\Omega=0)$} 
\put(-4.5,.8){\small $\Omega=\sqrt{2}\omega_\bot$} 
\put(-3.5,2.8){\small $\Omega=0$} 
\renewcommand{\figurename}{FIG.} 
\caption{\small Approach of the heat capacity (\ref{HC2}), solid lines, and its approximation (\ref{HC2DP}), 
dashed lines, to the Dulong-Petit law, horizontal line.} 
\label{PHC3} 
\end{minipage} 
\end{figure} 
\begin{eqnarray}
\label{HC2}
\frac{C_>}{k_BN}&=&\frac{35\zeta_4(e^{\beta\mu},\Omega)}{4\zeta_3(e^{\beta\mu},\Omega)}-\frac{25\zeta_3(e^{\beta\mu},\Omega)}{4\zeta_2(e^{\beta\mu},\Omega)}\nonumber\\[2mm]
&&\hspace{-15mm}+\gamma_T(\omega_\bot^2-\Omega^2)^2\left[\frac{11\zeta_3(e^{\beta\mu})}{2\zeta_3(e^{\beta\mu},\Omega)}-5\frac{\zeta_2(e^{\beta\mu})}{\zeta_2(e^{\beta\mu},\Omega)}\right]
\nonumber\\[2mm]
&&\hspace{-15mm}+\gamma_T^2(\omega_\bot^2-\Omega^2)^4\frac{\zeta_2(e^{\beta\mu})\zeta_2(e^{\beta\mu},\Omega)-\zeta_2^2(e^{\beta\mu})}{\zeta_2(e^{\beta\mu},\Omega)\zeta_3(e^{\beta\mu},\Omega)}\,.
\end{eqnarray}
To obtain this result, we have determined the derivative $(\partial \beta\mu/ \partial T)_{N,V}$ from (\ref{N1}). 
In the limit $k\downarrow 0$ of a harmonic trap, the heat capacity (\ref{HC2}) reduces to the well-known result
\begin{equation}
\label{HC2L}
\frac{C_>}{k_BN}=12\frac{\zeta_4(e^{\beta\mu},\Omega)}{\zeta_3(e^{\beta\mu},\Omega)} - 9\frac{\zeta_3(e^{\beta\mu},\Omega)}{\zeta_2(e^{\beta\mu},\Omega)}\,;\,k=0\,.
\end{equation}
Again, both cases $\Omega=\omega_\bot$ and $\Omega\to\infty$ yield the same analytic expression:
\begin{eqnarray}
\label{HC2rl}
\frac{C_>}{k_BN} &=& \frac{35}{4}\frac{ \zeta_{7/2}\left( e^{\beta\mu + 
\gamma_T(\omega_\bot^2-\Omega^2)^2} \right) }{ \zeta_{5/2}\left( e^{\beta \mu + \gamma_T(\omega_\bot^2-\Omega^2)^2} \right) }\nonumber\\[2mm] 
&&-\frac{25}{4}\frac{ \zeta_{5/2}\left( e^{\beta\mu+\gamma_T(\omega_\bot^2-\Omega^2)^2} \right) }{ \zeta_{3/2}\left( e^{\beta\mu+\gamma_T(\omega_\bot^2-\Omega^2)^2} \right) } \, .
\end{eqnarray}
Now, we investigate the heat capacity (\ref{HC2}) in the high temperature limit $T\to\infty$. 
To this end we use a large $T$-expansion of the generalized $\zeta$-function (\ref{Zeta1}): 
\setlength{\unitlength}{1cm}
\begin{figure}[t]
\center\begin{minipage}[tb]{.4\textwidth}
\center\includegraphics[scale=.65]{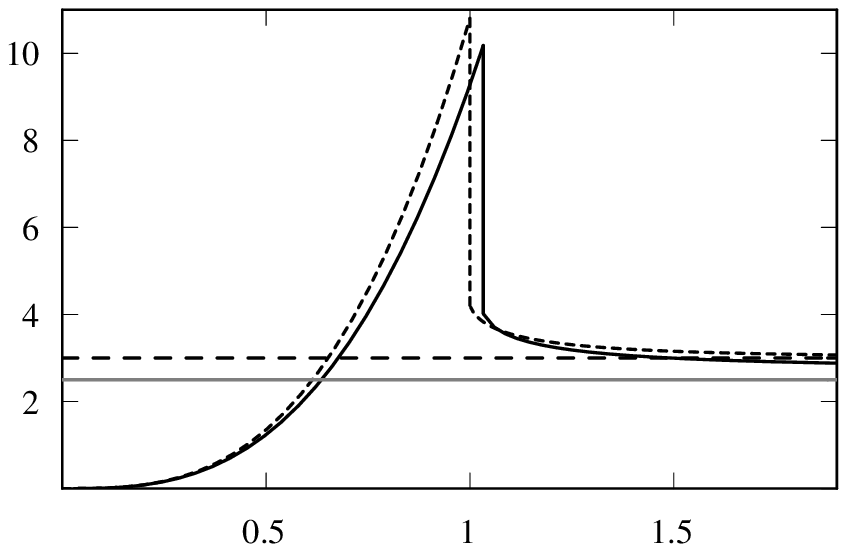}
\put(-4,-.3){$T/T_c(k=0)$}
\put(-7,1.5){\rotatebox{90}{$C/k_BN$}}
\put(-4.4,2.8){$k=0$}
\put(-2.9,2.3){$k\neq0$}
\renewcommand{\figurename}{FIG.}
\caption{\small Heat capacity versus temperature without rotation, reduced to the critical temperature (\ref{Tc0}) 
of the trap (\ref{Vrot1}) with $k=0$. The dashed line corresponds to the harmonic trap heat capacity (\ref{HC2L}), 
(\ref{HC3L}) for the Paris trap (\ref{Vrot1}) with $k=0$. The black solid line is the heat capacity (\ref{HC2}), 
(\ref{HC3}) for the anharmonic Paris trap \cite{Bretin}. The horizontal lines correspond to the Dulong-Petit law:  
harmonic trap (dashed) and anharmonic trap (gray).}
\label{PHC1}
\end{minipage}
\end{figure}
\begin{eqnarray}
\label{Zeta4}
\zeta_\nu\left( z,\Omega \right)\!\!&\approx&\! 
e^{\beta\mu}\left[ \sqrt{\pi\gamma_T}(\omega_\bot^2-\Omega^2) - 2\gamma_T(\omega_\bot^2-\Omega^2)^2 + ... \right]\nonumber\\
&&\hspace{-2cm} + \frac{e^{2\beta\mu}}{2^\nu}\!\left[\! \sqrt{2\gamma_T}(\omega_\bot^2-\Omega^2) - 4\gamma_T(\omega_\bot^2-\Omega^2)^2 \!+ ... \right]+ ...\,.
\end{eqnarray}
Inserting the expansion (\ref{Zeta4}) into the number of particles (\ref{N1}), we find for the first order of the 
fugacity $e^{\beta\mu}\approx2N\sqrt{\hbar^6\omega_z^2k}/\sqrt{\pi M^2k_B^5T^{5}}$, so that the heat capacity (\ref{HC2}) behaves like
\begin{equation}
\label{HC2DP}
\frac{C_>}{k_BN}\!\approx\!\frac{5}{2}+ \frac{\gamma_T}{4\,\sqrt{\pi}}(\omega_\bot^2-\Omega^2) - 
\frac{ \gamma_T^3 \left(4-\pi \right) }{ 32\,\pi^{3/2} }(\omega_\bot^2-\Omega^2)^3 \,. 
\end{equation}
Thus, the heat capacity approaches the Dulong-Petit law in an anharmonic trap $\lim_{T\to\infty}{C_>/(k_BN)}=5/2$ 
which is $1/2$ smaller than the corresponding one in the harmonic trap. 
Furthermore, the first $\Omega$ dependent term in (\ref{HC2DP}) changes its behavior, from being larger ($\Omega<\omega_\bot$) than 
the limit to being smaller ($\Omega>\omega_\bot$), see Fig.~\ref{PHC3}. \\
At the critical point, the harmonic heat capacity (\ref{HC2L}) reduces for small rotation frequencies $\Omega<\omega_\bot$ to
\begin{equation}
\label{HCTC2}
\lim_{T\downarrow T_c}{\frac{C_>}{k_BN}} = 12\frac{\zeta(4)}{\zeta(3)}-9\frac{\zeta(3)}{\zeta(2)}\approx4.23\,;\,\,k=0\,.
\end{equation}
In both limits $\Omega=\omega_\bot$ and $\Omega\to\infty$, the heat capacity (\ref{HC2rl}) at the critical point is given by
\begin{eqnarray}
\label{HCTC1}
\lim_{T\downarrow T_c}{\frac{C_>}{k_BN}} &=& \frac{35}{4}\frac{\zeta(7/2)}{\zeta(5/2)}-\frac{25}{4}\frac{\zeta(5/2)}{\zeta(3/2)} \approx 4.14\,.
\end{eqnarray}
\subsection{Condensate Phase}
Now we turn to the condensate phase $T<T_c$ where the chemical potential is given by (\ref{Mu1}). 
For the entropy, we obtain 
\begin{eqnarray}
\label{S2}
\frac{S_<}{k_BN}&=&\left(\frac{T}{T_c}\right)^3 \left\{ \frac{7\zeta_4(e^{\beta\mu_c},\Omega)}{2\zeta_3(e^{\beta_c\mu_c},\Omega)}
-\beta\mu_c\frac{\zeta_3(e^{\beta\mu_c},\Omega)}{\zeta_3(e^{\beta_c\mu_c},\Omega)} \right.\nonumber\\[2mm]
&&\hspace{-10mm}\left.+\gamma_T(\omega_\bot^2-\Omega^2)^2 \left[\frac{\zeta_3(e^{\beta\mu_c})-\zeta_3(e^{\beta\mu_c},\Omega)}{\zeta_3(e^{\beta_c\mu_c},\Omega)}\right] \right\}.
\end{eqnarray}
The internal energy below the critical temperature reads
\begin{eqnarray}
\label{U2}
\frac{U_<}{Nk_BT}&=&\left(\frac{T}{T_c}\right)^3\left\{\frac{5\zeta_4(e^{\beta\mu_c},\Omega)}{2\zeta_3(e^{\beta_c\mu_c},\Omega)}+\gamma_T(\omega_\bot^2-\Omega^2)^2\right.
\nonumber\\[2mm]
&&\hspace{-2cm}\left.\times\left[\frac{\zeta_3(e^{\beta\mu_c})-\Theta(\omega_\bot-\Omega)\zeta_3(e^{\beta\mu_c},\Omega)}{\zeta_3(e^{\beta_c\mu_c},\Omega)} \right]\right\}+\mu_c N\,,
\end{eqnarray}
where $\Theta$ denotes the Heaviside function.
Thus, we find for the heat capacity below the critical temperature with help of Eq.~(\ref{Mu1})
\begin{eqnarray}
\label{HC3}
\frac{C_<}{k_BN} &=& \left( \frac{T}{T_c} \right)^3 \left\{ \frac{35\zeta_4(e^{\beta\mu_c},\Omega)}{4\zeta_3(e^{\beta_c\mu_c},\Omega)} 
+ \gamma_T(\omega_\bot^2-\Omega^2)^2 \right.\nonumber\\
&&\nonumber\\
&&\hspace*{-7mm}\times \left[ \frac{ 11\zeta_3(e^{\beta\mu_c}) }{ 2\zeta_3(e^{\beta_c\mu_c},\Omega)} - 5\Theta(\omega_\bot-\Omega)
\frac{ \zeta_3(e^{\beta\mu_c},\Omega) }{ \zeta_3(e^{\beta_c\mu_c},\Omega) }\right]\nonumber\\[2mm]
&&\hspace*{-7mm}+\,\gamma_T^2(\omega_\bot^2-\Omega^2)^4\left[\Theta(\Omega-\omega_\bot)\frac{\zeta_2(e^{\beta\mu_c})}{\zeta_3(e^{\beta_c\mu_c},\Omega)}\right.\nonumber\\[2mm]
&&\hspace*{-7mm}-\left.\left.\Theta(\omega_\bot-\Omega)\frac{\zeta_2(e^{\beta\mu_c})-\zeta_2(e^{\beta\mu_c},\Omega)}{\zeta_3(e^{\beta_c\mu_c},\Omega)} \right]\right\}\,.
\end{eqnarray}
In the limit $k\downarrow 0$, Eq.~(\ref{HC3}) simplifies to
\begin{equation}
\label{HC3L}
\frac{C_<}{k_BN}=12\frac{\zeta(4)}{\zeta(3)}\left(\frac{T}{T_c}\right)^3\,;\,\,k=0\,.
\end{equation}
Thus, at the critical point it has the value
\begin{equation}
\label{HCTC3}
\lim_{T\uparrow T_c}{\frac{C_<}{k_BN}} = 12\frac{\zeta(4)}{\zeta(3)}\approx10.80\,;\,\,k=0\,.
\end{equation}
In both limits $\Omega=\omega_\bot$ and $\Omega\to\infty$, the heat capacity is given by
\setlength{\unitlength}{1cm} 
\begin{figure}[t] 
\center\begin{minipage}[t]{.4\textwidth} 
\center\includegraphics[scale=.65]{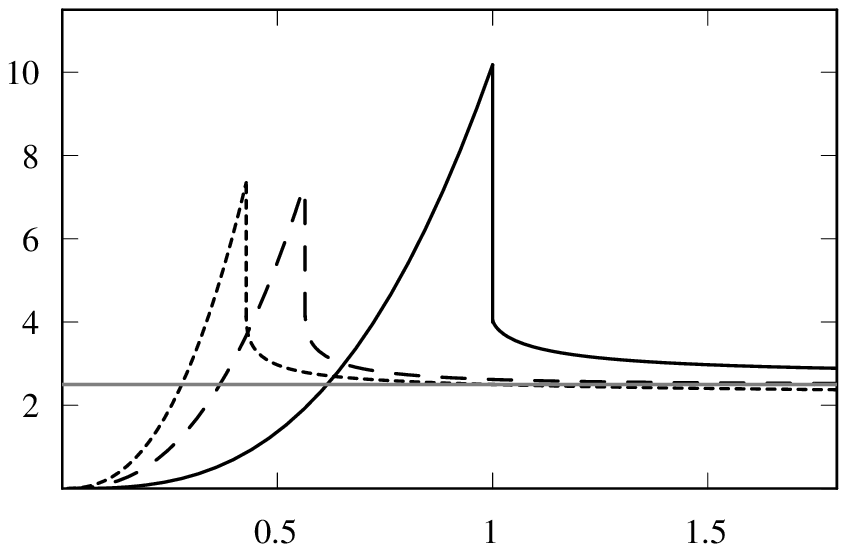} 
\put(-7,2){\rotatebox{90}{\small $C/k_BN$}} 
\put(-4,-0.3){\small $T/T_c(\Omega=0)$} 
\renewcommand{\figurename}{FIG.} 
\caption[Heat capacity just below  the critical point versus rotation frequency.]{\small Heat capacity versus 
temperature, reduced to the critical temperature $T_c(\Omega=0)$ of the anharmonic trap (\ref{Vrot1}) for varying 
rotation frequencies. The lines correspond to $\Omega=0$ (solid), $\Omega=\omega_\bot$ (long dashes), and $\Omega=2\omega_\bot$ (short dashes). 
The gray solid line corresponds to the Dulong-Petit law, the first term of (\ref{HC2DP}). We note that for $\Omega=2\omega_\bot$ 
and $T>T_c$ the heat capacity approaches the Dulong-Petit limit from below (see Fig.~\ref{PHC3}).} 
\label{PHC2} 
\end{minipage} 
\end{figure} 
\begin{equation}
\label{HC3lr}
\frac{C_<}{k_BN} = \frac{35}{4} \frac{\zeta(7/2)}{\zeta(5/2)}\left(\frac{T}{T_c}\right)^{5/2}
\end{equation}
so that the heat capacity (\ref{HC3}) at the critical point reduces to 
\begin{equation}
\label{HCTC4}
\lim_{T\uparrow T_c}{\frac{C_{<}}{k_BN}} = \frac{35}{4}\frac{\zeta(7/2)}{\zeta(5/2)}\approx7.35\,.
\end{equation}
In the low temperature limit $T\downarrow0$, the heat capacity (\ref{HC3}) tends to zero in accordance with the third law of 
thermodynamics. 
We note that the low-temperature limit of the heat capacity (\ref{HCLT}) has the same power-law behavior as the 
corresponding one of the condensate fraction (\ref{CF3}):
\begin{equation}
\label{HCLT}
\frac{C_<}{k_BN} \approx  \left\{
\begin{array}{ll}
{\displaystyle \frac{35k_B^3\zeta(4)}{4N\hbar^3\omega_z(\omega_\bot^2-\Omega^2)}\,T^3} &\!\!;\,\Omega<\omega_\bot\\[5mm]
{\displaystyle \frac{35M\sqrt{\pi}k_B^{5/2}\zeta(7/2)}{4N\sqrt{k}\hbar^3\omega_z}\,T^{5/2} } & \!\!;\,\Omega>\omega_\bot\,.
\end{array}\right.
\end{equation}
Fig.~\ref{PHC1} shows the temperature dependence of the heat capacity for the values of the Paris experiment without 
rotation \cite{Bretin}. 
We see that the effect of the anharmonicity is rather small. 
According to Ehrenfest's classification, the discontinuity at the critical temperature characterizes the phase 
transition to be of second order.
In Fig.~\ref{PHC2} we show how the heat capacity depends on the rotation frequency $\Omega$. 
Here, the rotation has a huge influence on the temperature dependence of the heat capacity. 
\section{Conclusions}
We have determined a semiclassical approximation for 
the critical temperature $T_c$ at which the condensation of a rotating  ideal Bose gas occurs in 
the anharmonic trap (\ref{Vrot1}). 
We have found that condensation is possible even in the overcritical rotation regime $\Omega>\omega_\bot$ which is in contrast with 
the harmonic trap where the condensate gets lost, when the rotation frequency $\Omega$ gets close to the trap frequency $\omega_\bot$. 
Our value for the critical temperature $T_c\approx64$~nK at the critical rotation frequency $\Omega=\omega_\bot$ 
yields $k_B T_c / \hbar \omega_z \approx 120 \gg 1$ which justifies a posteriori our semiclassical treatment. However, 
the value $T_c\approx64$~nK is about three 
times smaller than the one estimated in the Paris experiment \cite{Bretin}. 
This huge discrepancy could not be explained with the circumstance that our semiclassical analysis of the rotating ideal 
Bose gas neglects three important aspects, namely finite-size corrections, interactions between the particles, and the 
effect of vortices in the condensate. 
All three should have the effect of lowering the critical temperature. 
In a harmonic trap, it was found numerically that the finiteness of the system slightly lowers the critical 
temperature \cite{Ketterle96} which was also shown analytically \cite{Grossmann95,Haugset97}. 
Furthermore, an additional weak repulsive two-particle contact interaction leads to a negative shift of just a few 
percent \cite{Gio96,As04}. Therefore, we conclude that our semiclassical findings ask for a revised
experimental measurement of the critical temperature for a rotating Bose gas.\\
For the condensate fraction, we can state that the low-temperature behavior crucially depends on the rotation frequency 
$\Omega$. 
It shows a non-uniform temperature dependence which is in between the two power-laws $T^{5/2}$ and $T^3$. \\
The heat capacity of the rotating ideal Bose gas is discontinuous at the critical temperature. 
It tends in agreement with the third law of thermodynamics to zero in the low temperature limit $T\downarrow0$, and approaches the 
Dulong-Petit law (\ref{HC2DP}) in the high temperature limit $T\to\infty$ from above ($\Omega\leq\omega_\bot$) or from below ($\Omega>\omega_\bot$). \\
\section{Acknowledgement}
We thank Jean Dalibard, Hagen Kleinert, and Sabine Stock for critical reading of the manuscript as well as Robert Graham 
for the hospitality at the University Duisburg-Essen. 
Furthermore, support from the DFG Priority Program SPP 1116 {\it Interaction in Ultracold Gases of Atoms and Molecules} 
is acknowledged. 
\end{document}